\documentclass[aps,prd,preprint,superscriptaddress,tightenlines,%
nofootinbib]{revtex4}
\newcommand{\PRE}[1]{{#1}}   

\usepackage{bm}
\usepackage{epsfig}

\newcommand{\postscript}[2]{\setlength{\epsfxsize}{#2\hsize}
   \centerline{\epsfbox{#1}}}

\newcommand{\mbh}{M_{\text{BH}}}

\newcommand{\tev}{\text{TeV}}

\newcommand{\g}{\text{g}}

\newcommand{\etal}{{\em et al.}}

\newcommand{\eqref}[1]{Eq.~(\ref{#1})}

\def\sh{\sqrt{\hat s}}

\begin{document}

\title{
\PRE{\vspace*{.9in}}
Quest for Black Holes and Superstring Excitations \\ in Cosmic Ray Data
\PRE{\vspace*{0.3in}}
}

\author{Luis A.~Anchordoqui}
\affiliation{Department of Physics,\\
Northeastern University, Boston, MA 02115
\PRE{\vspace*{.1in}}
}

\author{Jonathan L.~Feng}
\affiliation{Department of Physics and Astronomy,\\
University of California, Irvine, CA 92697
\PRE{\vspace*{.1in}}
}

\author{Haim Goldberg}
\affiliation{Department of Physics,\\
Northeastern University, Boston, MA 02115
\PRE{\vspace*{.1in}}
}
\affiliation{Center for Theoretical Physics,\\ Massachusetts Institute of Technology, Cambridge, MA 02139
\PRE{\vspace*{.1in}}
}
\author{Alfred D.~Shapere}%
\affiliation{Department of Physics,\\
University of Kentucky, Lexington, KY 40506
\PRE{\vspace*{.5in}}
}

\begin{abstract}
\PRE{\vspace*{.1in}}
\noindent In this talk we discuss aspects of  TeV-scale gravitational
collapse to black holes and string balls and their subsequent evaporation.
Special emphasis is placed on the interplay of the
string $\rightleftharpoons$ black hole correspondence principle.
These ideas are then explored in the context of cosmic ray physics.
First, the potential for observing showers mediated by black
holes or superstring excitations is examined. Next, existing data from
neutrino telescopes are used to constrain the parameter space for the
unseen dimensions of the universe. Finally, we close with a discussion of
future prospects.\\

\PRE{\vspace*{.9in}}

\begin{center}
\framebox[1.1\width]{{\Large{\bf COSMO -- 03}}}
\end{center}

\begin{center}
{\tt http://www.ippp.dur.ac.uk/cosmo03/}
\end{center}

\end{abstract}


\maketitle

\section{Introduction}

A promising route towards reconciling the apparent mismatch of the
fundamental scales of
particle physics and gravity is to modify the short distance behavior of
gravity at scales much larger
than the Planck length. This can be accomplished in a straightforward
manner~\cite{Arkani-Hamed:1998rs,Antoniadis:1998ig} if one assumes that the
standard model (SM)
fields are confined to a 4-dimensional world (corresponding to our apparent
universe), while gravity lives
in a higher dimensional space. One virtue of this assumption is that very
large extra dimensions are
allowed without conflicting with current experimental
bounds~\cite{Hewett:2002hv}, leading to a
fundamental Planck mass much lower than its effective 4-dimensional value,
even as low as the electroweak scale. In particular, if the spacetime is 
taken as a direct product of a
4-dimensional spacetime and a flat spatial torus $T^{D-4}$
(of common linear size
$2\pi r_c$), one obtains a
definite representation of this picture in which the effective 4-dimensional
Planck scale,
$M_{\rm Pl} \sim 10^{19}$~GeV, is related to the fundamental scale of
gravity, $M_D$, according to $M^2_{\rm Pl} = 8 \pi \, M_D^{D-2} \,
r_c^{D-4}$~\cite{Arkani-Hamed:1998rs}.

One of the most startling predictions of scenarios of this sort is 
that microscopic
black holes (BHs) could be observed in forthcoming 
colliders~\cite{Banks:1999gd} and at existing
cosmic ray detectors~\cite{Feng:2001ib}.  In fact, the non-observation of 
cosmic neutrinos has already been used to set lower bounds on the fundamental
scale of $D$-dimensional gravity~\cite{Anchordoqui:2003jr}.  This work 
relied on classical
models of BH production, which are valid only for sufficiently
massive BHs, $M_{\rm BH} \gg M_D$.  For masses close to $M_D$,
gravity becomes strong and the classical description can no longer be
trusted.  String theory provides the best hope for understanding the regime
of strong quantum gravity, and in particular for computing cross sections at
energies close to the Planck scale~\cite{Dimopoulos:2001qe}. In principle 
embedding TeV-scale
gravity models in realistic string models might facilitate the calculation
of cross sections for BHs (and string excitations) of masses close to
$M_D$.

Our purpose here is to explore this possibility within a simple string
theory. We begin by reviewing the standard classical and semiclassical
pictures of BH production and decay, and the string $\rightleftharpoons$ 
BH correspondence
principle. We then calculate the $\nu N$ cross section for production of 
excited string states within superstring theory, and find it to be larger
than the BH production cross section~\cite{Cheung:2002aq}. This makes it 
plausible that the BH
cross section can be used as a lower bound down to $M_{\rm BH}=M_D$, 
which  would
significantly strengthen existing lower bounds on $M_D$.

To be specific, we will consider embedding of a 10-dimensional 
low-energy scale gravity scenario within the 
context of SO(32) Type I superstring theory,
where gauge and charged SM fields can be identified with open strings
localized on a 3-brane and the gravitational sector consists of
closed strings that propagate freely into the internal dimensions of the
universe~\cite{Antoniadis:1998ig}. After compactification on $T^6$ down to
four dimensions, $M_{\rm Pl}$ is related to the string scale, $M_s$, and the
string coupling constant, $g_s$, by $M_{\rm Pl}^2 =
(2 \pi \,r_c)^6 \,M_s^8/g_s^2.$ Hereafter, $D=10$. Generalization to arbitrary 
number of dimensions is straightforward.

\section{Black Holes and String Balls with Low-Scale Gravity}

Analytic and numerical studies have revealed that gravitational 
collapse takes place at sufficiently high energies and small impact 
parameters, as conjectured years ago by Thorne~\cite{Thorne:ji}. In 
the case of  4-dimensional
head-on collisions~\cite{D'Eath:hb}, as well as those with
non-zero impact
parameter~\cite{Eardley:2002re}, a horizon forms when and
only when a mass is compacted into a hoop whose circumference in every
direction is less than $2 \pi$ times its Schwarzschild radius up to a factor 
of order 1. In
the 10-dimensional scenario the Schwarzschild radius still characterizes
the maximum impact parameter for horizon formation~\cite{Yoshino:2001ik}. In 
the course of collapse, a certain  amount of
energy is radiated in gravitational waves by the
multipole moments of the incoming shock waves~\cite{D'Eath:hb}, leaving
a fraction $y \equiv \mbh/\sh$ available to Hawking
evaporate~\cite{Hawking:1975sw}. Here, $\mbh$ is a {\it lower bound}
on the final mass of the BH and $\sh$ is the center-of-mass energy of
the colliding particles, taken as partons. This ratio depends on the
impact parameter of the collision, as well as on the dimensionality
of space-time~\cite{Yoshino:2002br}.

Subsequent to formation, the BH proceeds to decay~\cite{Chamblin:2003wg}.
The emission rate per degree of particle freedom
$i$ of particles of spin $s$ with initial total energy between $(Q, Q+dQ)$
is found to be~\cite{Han:2002yy}
\begin{equation}
\frac{d\dot{N}_i}{dQ} = \frac{\sigma_s(Q,r_s)\,\,\Omega_{d-3}}{(d-2)\,(2\pi)^{d-1}}\,\,Q^{d-2} \left[
\exp \left( \frac{Q}{T_{\rm BH}} \right) - (-1)^{2s} \right]^{-1} \,\,,
\label{rate}
\end{equation}
where $T_{\rm BH} = 7/(4\,\pi\,r_s)$ is the BH temperature,
\begin{equation}
\label{schwarz}
r_s (M_{\rm BH}) = \frac{1}{M_{10}} \left[ \frac{\mbh}{M_{10}} \,\,
8 \,\pi^{3/2}\,\, \Gamma\left({9/ 2}\right)
\right]^{1/7}\,
\end{equation}
is the Schwarzschild radius~\cite{Myers:un}, 
\begin{equation}
\Omega_{d-3} = \frac{2\,\pi^{(d-2)/2}}{\Gamma[(d-2)/2]}
\end{equation}
is the volume of a unit $(d-3)$-sphere, and $\sigma_s
(Q,r_s)$ is the absorption coefficient (a.k.a. the greybody
factor). Recall that SM fields live on a 3-brane ($d=4$), while 
gravitons inhabit the entire spacetime ($d=10$). The prevalent energies 
of the decay quanta 
are $\sim T_{\rm BH} \sim 1/r_s$, resulting in $s$-wave dominance of the
final state. Indeed, as the total angular momentum number of the
emitted field increases, $\sigma_s (Q,r_s)$ rapidly gets
suppressed~\cite{Kanti:2002nr}. In the low energy limit, $Q
\, r_s \ll 1,$ higher order terms get suppressed by a factor of
$3 (Q\,r_s)^{-2}$ for fermions and $25 (Q\,r_s)^{-2}$ for gauge
bosons. For an average particle energy, $\langle Q \rangle
\approx r_s^{-1},$ higher partial waves also get suppressed,
though the suppression is not as large. This strongly suggests that
the BH is only sensitive to the radial coordinate and does not
make use of the extra angular modes available in the internal
space~\cite{Emparan:2000rs}. A recent  numerical study~\cite{Harris:2003eg}
has explicitly shown that the emission
of scalar modes into the bulk is largely suppressed with respect
to the brane emission. In order to contravene the argument of
Emparan--Horowitz--Myers~\cite{Emparan:2000rs},
the bulk emission of gravitons would need to show the opposite
behavior -- a substantial enhancement into bulk modes. There is
no {\it a priori} reason to suspect this qualitative difference between
$s=0$ and $s=2$, and hence no reason to support
arguments~\cite{Cavaglia:2003hg} favoring  deviation from  the dominance
of visible decay. With this in mind,  we assume the evaporation
process to be dominated by the large number of SM
brane modes.

The total number of particles emitted is roughly equal to the BH entropy,
\begin{equation}
S_{\rm BH} = \frac{\pi}{2}\,\mbh\,r_s.
\end{equation}
At a given time, the rate of decrease in the BH mass is just
the total power radiated
\begin{equation}
\frac{d\dot{M}_{\rm BH}}{dQ} = - \sum_{i} c_i\, \frac{\sigma_s(Q, r_s)}{8 \,\pi^2}\,\,Q^3 \left[
\exp \left( \frac{Q}{T_{\rm BH}} \right) - (-1)^{2s} \right]^{-1}\,\, ,
\label{rate2}
\end{equation}
where $c_i$ is the number of internal degrees of freedom of particle
species $i$. Integration of
Eq.~(\ref{rate2}) leads to
\begin{equation}
\dot{M}_{\rm BH} = - \sum_i c_i \,\,f\,\, \frac{\Gamma_s}{8\,\pi^2} \,\, \,\Gamma(4) \,\,
\zeta(4)\, \,T^4_{\rm BH}\,A_4,
\label{m}
\end{equation}
where $f=1$ ($f = 7/8$) for bosons (fermions), and  the
greybody factor was conveniently written
as a dimensionless constant, $\Gamma_s = \sigma_s(\langle Q \rangle,r_s)/A_4$, normalized to the BH surface area~\cite{Emparan:2000rs}
\begin{equation}
A_4 = \frac{36}{7}\,\pi\,\left( \frac{9}{2} \right)^{2/7}\ \, r_s^2
\label{area}
\end{equation}
seen by the SM fields ($\Gamma_{s=1/2} \approx 0.33$ and
$\Gamma_{s=1} \approx 0.34$~\cite{greybody}). Now, since the ratio of
degrees of freedom for gauge bosons, quarks and leptons is
29:72:18 (the Higgs boson is not included), from  Eq.~(\ref{m}) one obtains a rough estimate of the mean lifetime,
\begin{equation}
\tau_{_{\rm BH}} \approx  1.67 \times 10^{-27}
\left(\frac{\mbh}{M_{10}}\right)^{9/7} \left(\frac{\rm TeV}{M_{10}}\right)
\,{\rm s}\,,
\label{lifetime}
\end{equation}
which indicates that BHs evaporate instantaneously into visible quanta.
This is within a factor of 2 of the heuristic
estimate made in Ref.~\cite{Argyres:1998qn},
\begin{equation}
\tau_{_{\rm BH}} \sim 6.58 \times 10^{-28}
\left(\frac{\mbh}{M_{10}}\right)^{9/7} \left(\frac{\rm TeV}{M_{10}}\right)
\,{\rm s}\,,
\end{equation}
which is generally used in the literature.

The semiclassical description outlined above is only reliable when
the energy of the emitted particle is small compared to the BH
mass, i.e.,
\begin{equation}
T_{\rm BH} \ll  M_{\rm BH}\,, \,\,\, {\rm or}\,\,\, {\rm equivalently,} \,\,\, \mbh \gg M_{10} \,,
\label{condition}
\end{equation}
because it is only under this condition that both the gravitational
field of the brane and the back reaction of the metric during the emission
process can be safely neglected~\cite{Preskill:1991tb}. For BH with initial
masses well above $M_{10}$, most
of the decay process can be well described within the semiclassical approximation. However,
the condition stated in Eq.~(\ref{condition}) inevitably breaks down during the last stages
of evaporation. At this point it becomes necessary to introduce
quantum considerations. To this end we turn to a quantum statistical
description of highly excited strings.

It is  well-known that the density of string states with mass
between $M$ and $M+dM$ cannot increase any faster than
$\rho (M) = e^{\beta_H M}/M,$ because the partition function,
\begin{equation}
Z (\beta) = \int_0^\infty dM\, \rho(M) \,\,e^{-M\, \beta} \,\,,
\end{equation}
would fail to converge~\cite{Hagedorn:st}. Indeed, the partition function
converges only if the temperature is less than the Hagedorn
temperature, $\beta_H^{-1}$, which is expected to be $\sim M_s$. As
$\beta$ decreases to the transition point $\beta_H$,
the heat capacity rises to infinity because the energy goes into
the many new available modes rather than into raising the kinetic
energy of the existing particles~\cite{Frautschi:1971ij}. In the
limit, the total probability diverges,
indicating that the canonical ensemble is inadequate for the
treatment of the systems. However, one can still employ a microcanonical
ensemble of a large number of similar insulated system each with a given
fixed energy $E$. With the center-of-mass at rest, $E = M$. This means that
the density of states is just $\rho(M)$ and the entropy
$S = \ln \rho(M)$. In this picture the equilibrium
among systems is governed by the equality of the temperatures, defined for each system as
\begin{equation}
T \equiv \left(\frac{\partial S}{ \partial M} \right)^{-1} = \frac{M}{\beta_H M -1} \ .
\end{equation}
Equilibrium is achieved at maximum entropy when the total system heat
capacity, $C$, is positive.  Ordinary systems (on which our intuition is
founded) have $C>0$. However, for a gas of massive superstring excitations
the heat capacity,
\begin{equation}
C \equiv -\frac{1}{T^2} \left(\frac{\partial^2 S}{\partial M^2}\right)^{-1} = - \left(\frac{M}{T}\right)^2\, ,
\end{equation}
is negative, as is the case for BHs~\cite{Hawking:de}.
The positivity requirement on the total specific heat implies that strings
and BHs cannot coexist in thermal equilibrium, because any subsystem of this
system has negative
specific heat, and thus the system as a whole is thermodynamically
unstable. This observation suggests that BHs may end
their Hawking evaporation process by making a transition to an excited
string state with higher entropy, avoiding the singular zero-mass
limit~\cite{Bowick:1985af}. The suggestion of a string $\rightleftharpoons$ BH transition is further strengthened
by three other facts: ({\it i}) in string
theory, the fundamental string length should set a minimal radius for the
Schwarzschild radius of any BH~\cite{Veneziano:1986zf}; ({\it ii})
$T_{\rm BH} \sim \beta_H^{-1}$ for $r_s \sim M_s^{-1}$~\cite{Susskind:ws}.
({\it iii}) There is a seeming correlation between the greybody factors in BH decay and the level structure of
excited strings~\cite{Das:1996wn}.
The string $\rightleftharpoons$ BH ``correspondence principle''~\cite{Horowitz:1996nw} unifies these concepts: When the size of
the BH horizon drops below the size of the fundamental string length $\ell_s \gg \ell_{10},$ where $\ell_{10}$ is the
fundamental Planck length, an adiabatic
transition occurs to an excited string state. Subsequently, the string
will slowly
lose mass by radiating massless particles with a nearly thermal spectrum at
the unchanging Hagedorn temperature~\cite{Amati:1999fv}.(The probability that
a BH will radiate a large string, or else that a large string would undergo
fluctuation to become a BH is very small~\cite{Horowitz:1997jc}.)

The continuity of the cross section at the correspondence point, at least
parametrically in energy and string coupling, provides independent supportive
argument for this picture~\cite{Dimopoulos:2001qe}.  Specifically,  (in the
perturbative regime) the Virasoro-Shapiro amplitude leads to
a ``string ball'' (SB) production cross section
$\propto g_s^2 \hat{s}/M_s^4$. This cross section saturates the unitarity
bounds at $g_s^2 \hat{s}/M_s^2 \sim 1$~\cite{Amati:1987wq}, so before
matching the geometric BH cross section $\propto r_s^2$ there is a transition
region at which $\hat{\sigma} \sim M_s^{-2}$.  All in all, the
rise with energy of the parton-parton $\rightarrow$ SB/BH cross section
can be parametrized as~\cite{Dimopoulos:2001qe}
\begin{eqnarray}
\hat \sigma (\sqrt{\hat{s}})\sim\left\{
\begin{array}{ll}
\displaystyle
{g_s^2 \,\hat{s}\over M_s^4} &\qquad M_s\ll \sqrt{\hat{s}} \leq M_s/ g_s\,,\\
\displaystyle
{1\over M_s^{2}}&\qquad M_s/ g_s< \sqrt{\hat{s}} \leq M_s/ g_s^2\,,\\
\displaystyle
{1\over M_{10}^{2}}\,\left[{\sqrt{\hat{s}} \over M_{10}}\right]^{2/7}
&\qquad M_s/ g_s^2<\sqrt{\hat{s}}\,,
\end{array}
\right.
\end{eqnarray}
where $M_{10} = (8 \pi^5)^{1/8}\,M_s/g_s^{1/4}.$

Colliders have not yet attained the energies required to probe
TeV scale gravitational collapse; nevertheless cosmic rays may already
provide some clues. To date a handful of cosmic rays have been observed with
energy in excess of $10^{11}$~GeV~\cite{Anchordoqui:2002hs}. When these
particles impinge on a
stationary nucleon in the upper atmosphere they  probe center-of-mass
energies as high as $\sqrt{s} \sim 400$~TeV. In what follows we estimate the
sensitivity of cosmic ray observatories to SB/BH production.

\section{Probes of large extra dimensions with cosmic rays}

The SB/BH production cross section, ${\cal O} (M_{\rm EW}^{-1})$, is about 5
orders of magnitude smaller than QCD cross sections, ${\cal O}
(\Lambda_{\rm QCD}^{-1}),$ thus making it futile to hunt for SB/BHs in
hadronic cosmic rays.  On the other hand,  SM neutrinos interact so weakly
that any quantum gravitational enhancement of the
cross section can be experimentally distinguished from background.
A substantial neutrino component may accompany the observed cosmic ray flux
because of neutrino production in cosmic beam dumps (such as gamma ray burst
fireballs~\cite{Razzaque:2002kb}, $X$-ray binaries~\cite{Anchordoqui:2002xu},
or active galactic nuclei~\cite{Szabo:qx}). In addition, the inelastic
collision of ultra-high energy~($> 10^{10.6}$~GeV) nucleons with the relic
photons permeating the universe should deplete the hadronic cosmic ray
intensity at the end of the spectrum and breed neutrinos from decay products
of charged pions~\cite{Stecker:1979ah}. The reaction generating these
cosmogenic
neutrinos is well known physics and the flux depends
only on the existence of ultra-high energy nucleon sources more distant
than~$\approx 8$~Mpc. In this case, the expected $\nu$-flux can be accurately
predicted by fitting the observed spectrum with a homogeneous
population of nucleon
sources~\cite{Protheroe:1996ft}.  The existence of the cosmogenic neutrinos
is a safe bet, implying a minimum guaranteed flux for testing any new
physics that promotes the neutrino cross section to a sub-hadronic
size.\footnote{In the simple $T^6$ compactification, there are some
unitarization procedures for the multi Kaluza-Klein graviton exchange
amplitudes which can yield hadronic cross sections at
$E_\nu \agt 10^{10}$~GeV~\cite{Jain:2000pu}. However, these cross sections have
quadratic energy dependence, and so will generate moderately
penetrating showers with definite profiles at lower
energies~\cite{Anchordoqui:2000uh}. These have not been reported to date, and
their absence places serious constraints on the enhancement of the
neutrino-nucleon cross section to sub-hadronic size.}

The inclusive production of BHs proceeds through different final
states for different classical impact
parameters $b$~\cite{Yoshino:2002br}. These final states are characterized
by the fraction $y(z)$ of the initial parton center-of-mass
energy, $\sqrt{\hat s}=\sqrt{xs}$, which is trapped within
the horizon. Here, $z= b/b_{\rm max},$ and $b_{\rm max}=
1.3 \, r_s(\sqrt{\hat s})$~\cite{Yoshino:2002br}.
With a lower cutoff $M_{\rm BH,min}$
on the BH mass required for the validity of the semi-classical description,
this implies a joint constraint
\begin{equation}
 y(z)\,\,\sqrt{x s} \ge M_{\rm BH,min}
\label{constraint}
\end{equation}
on the parameters $x$ and $z$. Because of the monotonically decreasing nature
of $y(z)$, Eq.~(\ref{constraint}) sets an {\it upper}
bound $\bar z(x)$ on the impact parameter for
fixed $x.$ The corresponding parton-parton BH cross section
is $\hat \sigma_{_{\rm BH}} (x) = \pi \bar b^2(x),$
where $\bar b=\bar z b_{\rm max}.$  The total BH production cross section
is then~\cite{Anchordoqui:2003jr}
\begin{equation}
\sigma_{_{\rm BH}}(E_\nu,M_{\rm BH,min},M_{10}) \equiv \int_{\frac{M_{\rm BH,min}^2}{
y^2(0) s}}^1 \, dx
\,\sum_i f_i(x,Q) \ \hat \sigma_{_{\rm BH}}(x) \,\,,
\label{sigma}
\end{equation}
where $i$ labels parton species and the $f_i(x,Q)$ are parton
distribution functions (pdfs)~\cite{Pumplin:2002vw}.
The momentum scale $Q$ is taken as $r_s^{-1},$ which is a typical
momentum transfer during the gravitational collapse. A useful criterion for
a BH description is that the entropy is
sizable, so that Eq.~(\ref{condition}) is satisfied.
For $M_{\rm BH,min} = 3\,M_{10}$,
$S_{\rm BH} \approx 13 \gg 1$.

In the perturbative string regime, i.e., $M_{\rm SB,min} <
\sqrt{\hat{s}} \leq M_s/g_s$, the SB production cross
section is taken as
\begin{equation}
\sigma_{_{\rm SB}}(E_\nu,M_{\rm SB,min}, M_{10}) = \int_{\frac{M_{\rm SB,min}^2}{s}}^1
dx   \,\sum_i f_i(x,Q) \, \hat\sigma_{_{\rm SB}}(\hat{s}) \,,
\end{equation}
where $\hat\sigma_{_{\rm SB}} (\hat{s})$ contains the Chan Paton factors
which control the projection of the initial state onto the string spectrum.
In general, this projection is not uniquely determined by the low-lying
particle spectrum,
yielding one or more arbitrary constants.
The analysis in the  $\nu q \rightarrow \nu q$
channel illustrates this point~\cite{Cornet:2001gy}. The $\nu g$
scattering, relevant for $\nu N$ interactions at ultra-high energies,
introduces additional ambiguities. For
simplicity, in the spirit of~\cite{Dimopoulos:2001qe}, we
take  $\hat{\sigma}_{_{\rm SB}} = (\pi^2 \g_s^2 \,\hat{s})/(8 M_s^4),$
derived from the Virasoro-Shapiro amplitude. The momentum scale in this
case varies from $Q\simeq \sqrt{\hat s}$
at parton center-of-mass energies corresponding to low-lying
string excitations, to
the BH value  $Q \approx r_s^{-1}$ for
$\sqrt{\hat s}\approx M_s/g_s^2,$ the transition point energy. This corresponds
to very high string excitations energies. In practice, this may be
approximated as a two stage designation:
\begin{eqnarray}
Q (\sqrt{\hat s}) \approx \left\{
\begin{array}{ll}
\displaystyle
\sqrt{\hat s}
&\qquad {\rm if}\,\,\sqrt{\hat{s}} \leq M_s/g_s^2  \,,\\ \\
\displaystyle
r_s^{-1}
&\qquad {\rm if}\,\, M_s/g_s^2 \,  < \sqrt{\hat{s}} \,,
\end{array}
\right.
\end{eqnarray}
with small ($\approx  10\%$) uncertainty~\cite{Anchordoqui:2002vb}.

\begin{figure}
\postscript{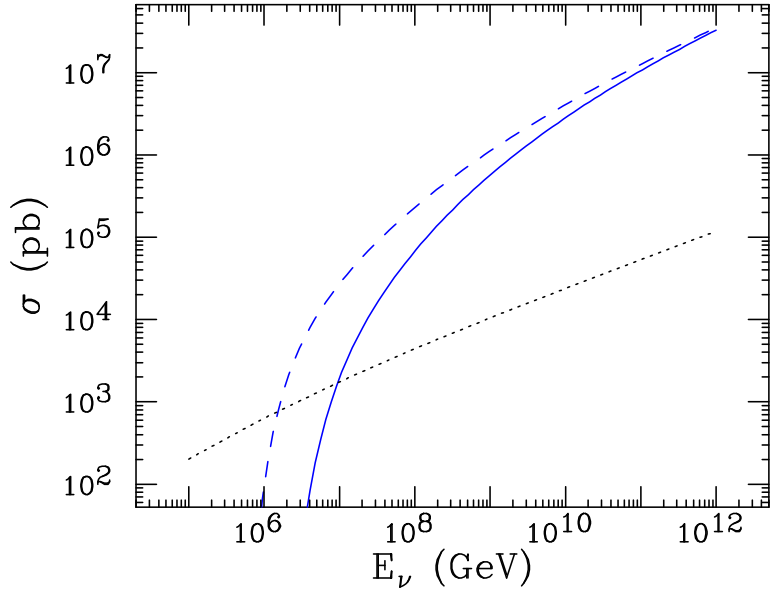}{0.65}
\caption{The solid line indicates the cross
section $\sigma(\nu N \to {\rm BH})$ for $M_{10} = 1~\tev$
and $M_{\rm BH,min}~=~M_{10}$. Energy loss has been included
according to \eqref{sigma}.  The dashed line indicates the SB production
cross section for $g_s = 0.2$. The very lowest threshold was set to
$M_{\rm SB,min} = 2 M_s$. The SM cross section
$\sigma(\nu N \to \ell X)$ is indicated by the dotted line.}
\label{sigma39}
\end{figure}

In Fig.~\ref{sigma39} we show a comparison between the total BH
production cross section and the particular model we are
considering for SBs. As can be seen in
Fig.~\ref{sigma39}, for $E_\nu \agt 10^7$~GeV, $\sigma_{_{\rm
SB}}> \sigma_{_{\rm
BH}} > \sigma_{_{\rm SM}}$. This makes plausible that in the quantum range,
$M_{10} < M_{\rm BH, min} < 3 M_{10},$ the BH cross can be thought
of as a lower bound on the anomalous $\nu N$ cross section. Since the BH 
production cross
section is suppressed by $M_{10}^2,$ bounds on the $\nu N$ cross
section can be translated into lower limits on $M_{10}.$
If the inequality above holds true in more realistic string models,
the bounds on $M_{10}$ can be significantly strengthened by taking
$M_{\rm BH,min}=M_{10}.$

Several techniques have been used to search for the cosmic neutrino
component~\cite{Halzen:2002pg}.
Showers in the atmosphere can be detected either by observing the fluorescence
or \v{C}erenkov light induced by the ionizing tracks in the air, or
by directly detecting the charged particles in the shower tail with
scintillation counters scattered on the Earth's surface.
Neutrino showers mediated by SB/BHs may be distinguished from hadronic
primary particles by looking for very inclined showers. This is because
for a horizontal path (90$^\circ$ zenith angle) the
neutrino interaction length for SB/BH production is larger than
the thickness of the atmosphere (36000 g/cm$^2$) and thus, unlike
hadrons that shower high in the atmosphere, neutrinos could travel through
most of this matter before  gravitational collapse is triggered. An alternative
technique exploits naturally occurring large volume \v{C}erenkov
radiators such as the Antarctic ice-cap.  Several radio antennae monitor cold
ice for radio frequency \v{C}erenkov radiation resulting from neutrino in-ice
cascades. Potential signal events are distinguished from background using
vertex location and signal shape.

The experimental situation is that, in spite of extensive searching, only one
neutrino-like event has been reported, with an expected background from
hadronic
cosmic rays of 1.72 events. This implies, at 95\% C.L., a maximum of 3.5
events, which combined with the sum of the exposures of the various
experiments, and convoluted
with the guaranteed cosmogenic neutrino flux, allows the
extraction of a lower bound on the fundamental Planck scale,
$M_{10} > 1.0 - 1.4$~TeV, for $M_{10} < M_{\rm BH,min} <
3 \,M_{10}$~\cite{Anchordoqui:2003jr}. This bound is competitive with that
obtained from Tevatron data~\cite{Abbott:2000zb}, and these represent the
most stringent lower limits to date for the size of 6 large extra dimensions.

\section{Concluding remarks}

We have reviewed the possibility of searching for black holes and superstring
excitations in cosmic rays using neutrino interactions in the Earth's
atmosphere and the Antarctic ice. The fact that neutrino showers have not yet
been observed allows us to put a lower bound on the fundamental Planck scale
$M_{10} > 1.0 - 1.4$~TeV. As discussed previously, the string
$\rightleftharpoons$ BH
correspondence principle provides support for the higher bound.
The next generation of cosmic ray
experiments~\cite{Anchordoqui:2001cg,Ringwald:2001vk} and
dedicated neutrino telescopes~\cite{Kowalski:2002gb} should
probe values of $M_{10}$ up to about 5~TeV.

Conversely, if neutrino showers are observed above predicted SM rate, they can be ascribed
to either new physics or a larger than expected cosmogenic flux.
The most powerful technique to distinguish between these two possibilities
exploits the absorption of SB/BH secondaries by the Earth,
by separately binning Earth skimming events~\cite{Feng:2001ue} which arrive
at very small angles to the horizontal.  An enhanced flux will
increase both quasi-horizontal and Earth-skimming event rates, whereas a
large SB/BH cross section suppresses the latter, because the hadronic decay
products of SB/BH evaporation do not escape the Earth's
crust~\cite{Anchordoqui:2001cg}. If such a suppression is observed then
string balls and black holes will provide a compelling explanation.

\begin{acknowledgments}
LAA would like to thank Sacha Davidson and the organizers of
COSMO-03 for a most stimulating meeting. The work of LAA and
HG has been supported in part by the US National Science Foundation (NSF),
under grants No.\ PHY--0140407 and No.\ PHY--0073034, respectively.
The work of JLF is supported in part by NSF CAREER grant No. PHY-0239817.
The work of ADS is supported in part by Department of Energy Grant
No.\ DE--FG01--00ER45832 and NSF Grants PHY--0071312 and PHY--0245214.
\end{acknowledgments}


\end{document}